\documentclass[12pt,preprint]{emulateapj}

\usepackage{natbib}

\usepackage{graphics,graphicx}

\newcommand{\Cha}{{\it Chandra~}}

\shorttitle{}

\shortauthors{Migliori G. et al.}

\begin{document}

\title{Broad-band Jet emission in young and powerful radio sources:

  the case of the CSS Quasar 3C~186 }

\author{Giulia Migliori}

\affil{Harvard-Smithsonian Center for Astrophysics, 60 Garden St., Cambridge, MA, 02138, USA}
\affil{SISSA, Via Bonomea 265, I-34136, Trieste, Italy}

\email{migliori@cfa.harvard.edu}

\author{Aneta Siemiginowska}

\affil{Harvard-Smithsonian Center for Astrophysics, 60 Garden St., Cambridge, MA, 02138, USA}

\author{Annalisa Celotti}

\affil{SISSA, Via Bonomea 265, I-34136, Trieste, Italy}

\begin{abstract}
\noindent
We present the X-ray analysis of a deep ($\sim$ 200 ksec) \Cha
observation of the compact steep spectrum radio-loud quasar 3C~186
(z=1.06) and investigate the contribution of the unresolved radio jet to the
total X-ray emission. The spectral analysis is not conclusive on the origin of the bulk of the X-ray emission. In order to examine the jet contribution to the X-ray flux,  
we model the quasar spectral energy distribution (SED), adopting several scenarios for the jet emission. For the values of the main physical parameters favored by the observables, 
a dominant role of the jet emission in the X-ray band is ruled out when a single zone (leptonic) scenario is adopted, even including the contribution of the external photon fields as seed photons for inverse Compton emission. We then consider a structured jet, 
with the blazar component that-  although not directly visible in the X-ray band - provides an intense field of  seed synchrotron photons Compton-scattered by  electrons in a mildly relativistic knot. In this case the whole X-ray emission can be accounted for if we assume a blazar luminosity within the range observed from flat spectrum radio quasars.
The X-ray radiative efficiency of such (structured) jet is intimately related to the presence of a complex velocity structure. The jet emission can provide a significant contribution in X-rays if it decelerates within the host galaxy, on kiloparsec scales. We discuss the implications of this model in terms of jet dynamics and interaction with the ambient medium.
\end{abstract}


\section{Introduction}
\noindent
Extragalactic jets have been observed on many physical scales and they
often extend to distances of hundreds of kiloparsec from the nucleus
\citep{Bri84,Zen97}.  Although radio jets have been known for several decades, only
 the \Cha X-ray Observatory observations have revealed their
 X-ray emission on large scales \citep[see for a review][] {HK06}.

 The broad-band emission of bright features observed in jets can be
 usually - though not uniquely - well accounted for by a two-component
 model: a low-energy synchrotron component and a high-energy one due
 to Compton (IC) up-scattering of seed photons off relativistic
 leptons.  The seed photons can be produced both `locally' in the jet
 \citep[synchrotron self Compton, SSC,][]{MGC92,BM96}, and externally.
 At hundred kpc scales, the X-ray emission of jets in powerful
 Fanaroff-Riley II (FRII) sources might be explained by the inverse
 Compton (IC) scattering of cosmic microwave background (CMB) photons
 \citep{Tav00,Cel01,Sie02,Sam04}.  At the parsec (and smaller) scales,
 the jet radiation could be due to IC scattering of photons originated
 in disk, Broad
 Emission Line Regions or hot dust \citep[see][]{DS93,SBR94,GM96,Bla04}.\\
 However, in radio-loud (RL) quasars the pc scale jets remain
 spatially unresolved at X--ray energies and often the high energy
 spectral information is limited by photon statistics and broad band
 coverage.  As a consequence, it is hard to disentangle the
 contribution of the non--thermal (and relativistically beamed)
 emission of the small-scale jet from the one related to the accretion
 processes (either the X-ray thermal emission of the innermost part of
 the disk or that resulting from Comptonization of disk photons by
 electrons in a hot
 corona).\\
 In the nuclear region, pieces of evidence for the contribution of
 X-ray emission from jets come from comparative studies of radio-quiet
 (RQ) and RL quasars, as RL quasars display a X-ray excess with
 respect to the RQ ones, which is most likely associated with the jet
 \citep[][and references therein]{Zam81,Wor87,Mil11}.  Because the jet
 emission is anisotropic, its relevance depends on the orientation of
 the jet axis with respect to the observer's line of sight. The
 presence of a non-thermal component is also supported by the fact
 that RL quasars are X-ray brighter (with respect to their optical
 luminosity) for increasing values of the radio loudness parameter,
 RL\footnote{The radio loudness parameter, RL, is defined as the
   logarithmic ratio of the monochromatic core fluxes at 5 GHz and
   2500 $\AA$ \citep{Kel89}.} \citep[see][]{Mil11}. \\

\noindent
 In young and compact radio sources, the situation is even more
 complex.  Due to their small linear sizes ($\lesssim$20 kpc), the
 entire radio structure of GigaHertz-Peaked and Compact Steep
 Spectrum\footnote{These radio sources are characterized by a linear
   size $\lesssim 20$ kpc and a power at 1.4 GHz $P_{1.4\,GHz}\geq
   10^{32}$ erg s$^{-1}$ Hz$^{-1}$. They typically display convex
     radio spectra with turn over frequencies between $\sim$0.1 and
     $\sim$1 GHz \citep[see][for a review]{odea98}.} (GPS, CSS) radio
 sources is typically enclosed in a region corresponding to the
 unresolved X-ray core of a giant quasar/radio source \citep{Sie09}.
 Thus, studies to understand the origin of their X-ray emission
 \citep{Gua06,Vin06,Sie08,Ten09} mainly rely on the analysis of the
 spectral features, which is hampered by the limited statistics;
 furthermore there is (yet) no information on possible $\gamma$-ray
 emission, which would indicate the contribution of non-thermal
 radiation from jets.  As a consequence, the origin of the bulk of the
 X-ray radiation in these objects is still unclear.  Understanding its
 origin in young sources is key to address several questions, such as
 how the source is interacting with the environment in its initial
 phase of expansion and how it will evolve. Furthermore estimating the
 bolometric jet emission could be a plausible indicator of the jet
 power and the fate of the radio source itself. \\
 From a statistical point of view, \citet{Ten09} found that GPS {\it
   galaxies} have an intrinsic X-ray luminosity comparable to FR II
 radio galaxies, where the bulk of the X-rays is generally related to
 the disk-corona system. The location of GPS galaxies in the X-ray to
 [OIII] luminosity ratio versus column density plane is also found to
 be coherent with this scenario. On the other hand, the authors point
 out that GPS galaxies seem to follow the same 2-10~keV and 5~GHz core
 luminosity correlation of FR~I radio galaxies \citep{Chia99,HW00},
 which points
 to a non-thermal, jet-related origin of the X-ray emission.\\
 A significant level of X-ray flux can be produced via IC of different
 seed photons in the compact lobes of GPS radio galaxies
 \citep{Sta08}. Given the typical GPS linear sizes ($\lesssim$1 kpc)
 the nuclear photon fields, e.g.  optical-UV disk and IR torus
 photons, are intense enough to provide, when Comptonized, X-ray
 luminosities of the order of $10^{44}$--$10^{46}$~erg~s$^{-1}$
 \citep{Sta08}. The model satisfactorily accounts for the properties
 of a sample of GPS galaxies with compact-symmetric-object morphology
 \citep{Ost10} whose emission is presumably not strongly beamed. For
 the case of a powerful (i.e. with a jet kinetic power
 $L_{kin}>10^{46}$ erg s$^{-1}$) and nearby ($\lesssim$1~Gpc) GPS
 source the model by Stawarz et al. (2008) also predicts $\gamma$-ray
 fluxes possibly detectable with the Large Area Telescope on board the
 Fermi satellite \citep[see also the model results for the case of
 4C~+55.17,][]{McC11}.\\
\noindent
 In the case of GPS and CSS {\it quasars}, the closer jet alignment to
 the line of sight should favor the beamed jet component. This is
 certainly observed in the radio band \citep{Fan90}, while the X-ray
 behavior remains more elusive.  The median X-ray photon index,
 $\langle \Gamma_X \rangle=1.84\pm0.24$, in the GPS and CSS quasar
 sample observed with \Cha \citep{Sie08} is larger than the typical
 index of RL quasars \citep[$\langle\Gamma_X
 \rangle=1.55\pm0.17$,][]{Bel06}. Also, the average optical-to-X-ray
 luminosity ratio\footnote{The optical-to-X-ray luminosity ratio,
   $\alpha_{OX}$ \citep{AT82}, is defined as $\log
   [F(2500\,\AA)/F(2\,keV) ]/2.605$ in rest frame, where $F(2500\
   \AA)$ and $F(2\ keV)$ are monochromatic fluxes in cgs units at 2500
   $\AA$ and 2 keV, respectively.}
 $\langle\alpha_{OX}\rangle=1.53\pm0.24$ agrees with the median value
 found for radio-quiet quasars
 \citep[$\langle\alpha_{OX}\rangle=1.49\pm0.19$,][]{Kel07,Sob09}
 pointing to an accretion related origin.  However, in the majority of
 cases we are still lacking the detection of characteristic features
 associated with the accretion flow, chiefly a Fe K$\alpha$
 fluorescent emission line associated with a reflection component
 suggesting that a jet X-ray component contaminates the observed
 spectrum.  Interestingly, the sources of the sample observed with
 \Cha appear to be more radio-loud than their giant counterparts
 \citep{Sie08}.  And indeed evolutionary models predict that the
 extended components of radio sources emit more efficiently during the
 initial phase of expansion \citep{Beg99}.  Furthermore, there are
 examples of CSS quasars whose X-ray spectrum displays a flat
 component in the hard X-rays, interpreted as
 emission from the base of the jet \citep[e.g. 3C~48,][]{Wor04}.\\
 
 \noindent
 In this context, we aim at studying the possible jet contribution in
 a sample of young and compact radio sources, for which we have a
 reasonable multi-band coverage of the SEDs, by modeling their
 broadband emission.  As a pilot case, we start our investigation from
 the case of 3C~186, a young RL quasar with a knotty radio jet that is
 unresolved in the \Cha image. Interestingly, 3C~186 is among the few
 quasars found in a X-ray cluster at ``high'' redshift (z=1.06).
 Hence, constraining whether the bulk of the quasar X-ray emission is
 thermal or non-thermal can also provide information on the jet power,
 in turn
 unveil the main channel through which the source interacts with the galactic and cluster environment.\\
 We model the jet SED as synchrotron and IC emission, taking into
 account the main local and external radiation fields which can act as
 seed photons.  As the assumption that a single (homogeneous) emitting
 region dominates the jet SED in quasars is still a matter of debate
 \citep[][and references therein]{Ghis09,Sik09,Pou10,Mars10,Lyu10}, we
 also treat the case of a jet with a complex velocity structure.  We
 consider the presence of two emitting regions moving with different
 velocities and taking into account the relative effects on the
 emission via IC. We also examine the implications of such scenario in
 terms of source dynamics and energetics.

\noindent
The paper is organized as follows. After reviewing the 3C~186
properties (\S \ref{sect:3C186}), in \S \ref{sect:Chandra} we focus on
the \Cha observations and the corresponding X-ray spectral analysis.
The model for the broadband emission from the quasar jet is described
in \S \ref{sect:Model}.  The results of the SED modeling and the
consequences in terms of jet power are discussed in \S
\ref{sect:Discussion}. We finally summarize our findings in \S
\ref{sect:Summary}.  Throughout this work, we assume the following
cosmological parameters H$_0$=71 km s$^{-1}$ Mpc$^{-1}$,
$\Omega_M$=0.27, $\Omega_\Lambda$=0.73.

\section{The radio-loud quasar 3C\,186}\label{sect:3C186}
\noindent
3C~186 is a luminous quasar \citep[$L_{bol}\sim10^{47}$ erg
s$^{-1}$][]{Sie05}, with a compact FRII morphology, located at
redshift $z=1.067$. The radio structure is characterized by two
components, identified as the radio lobes, separated by
$\sim$2$^{\prime\prime}$, and a knotty jet connecting the core to the
northwest lobe \citep[][]{Spe91}.\\
\noindent
The steep radio spectrum, the relatively small projected linear size 
($\sim16$ kpc), and an estimated spectral age of $\sim 10^{5}$ yrs \citep{Mur99}
indicate that the source belongs to the class of Compact Steep
Spectrum (CSS) radio quasars.  \\
At optical--UV wavelengths, 3C~186 has a typical quasar spectrum,
dominated by a luminous big blue bump component
\citep[$L_{UV}=5.7\times10^{46}$ erg s$^{-1}$,][]{Sie05}, and broad
optical emission lines \citep{Net96,SR00,Kur02,EK04}.\\
In the X-ray band, the quasar has been observed twice with the \Cha
satellite.  The first \Cha observation led to the discovery of
diffuse ($r\gtrsim500$ kpc) X-ray emission surrounding the quasar,
interpreted as thermal radiation from the intracluster medium
\citep{Sie05}.  This is one of the few cases of a quasar in a luminous
X-ray cluster environment at $z> 1$.  The quasar is bright in X-rays
($L_X(2-10\ \rm keV)=1.2\times 10^{45}$ erg s$^{-1}$) and dominates
over the thermal cluster emission in the central region \citep[the
X-ray diffuse emission is $\lesssim3\%$ of the total X-ray
flux,][]{Sie10}.  A detailed study of the second deep \Cha observation - dealing with the extended component - 
shows that the cluster gas temperature drops from $kT\sim8$ keV in the
outer parts of the cluster to $kT\sim3$ keV in the central region,
indicating a strong cooling core \citep{Sie10}. The authors argued
that the cooling gas could provide enough fuel to support the growth of
the central supermassive black hole. The quasar in turn would provide the
energy/momentum critical to the possible radiative feedback 
which is still not well understood.\\
The morphological study of the X-ray emission of the quasar is hampered by the small
angular dimensions of the radio source.  The X-ray spectral analysis based on the
first \Cha observation dataset was not conclusive on the nature and origin of the X-ray emission.  The best fit
model, a simple power-law with a steep spectral index
\citep[$\Gamma_X=2.01\pm0.07$,][]{Sie05}, is compatible with being both non-thermal emission from the extended radio components,
namely jet and lobes, as well as a nuclear emission, related to the
disk photons Comptonized by electrons in a hot corona.  
The presence of the Fe K$\alpha$ line in the quasar X-ray spectrum,
which would point to the nuclear origin hypothesis,  was
uncertain due to the limited statistics.  We start our study with a
description and analysis of the second \Cha observation of the quasar emission. We also use
the archival data from the first observation which we reprocessed in
order to apply the updated calibration.

\section{Chandra Observations}\label{sect:Chandra}
\subsection{Observations and Data Analysis}
\begin{figure*}
\centering
\includegraphics[scale=0.35]{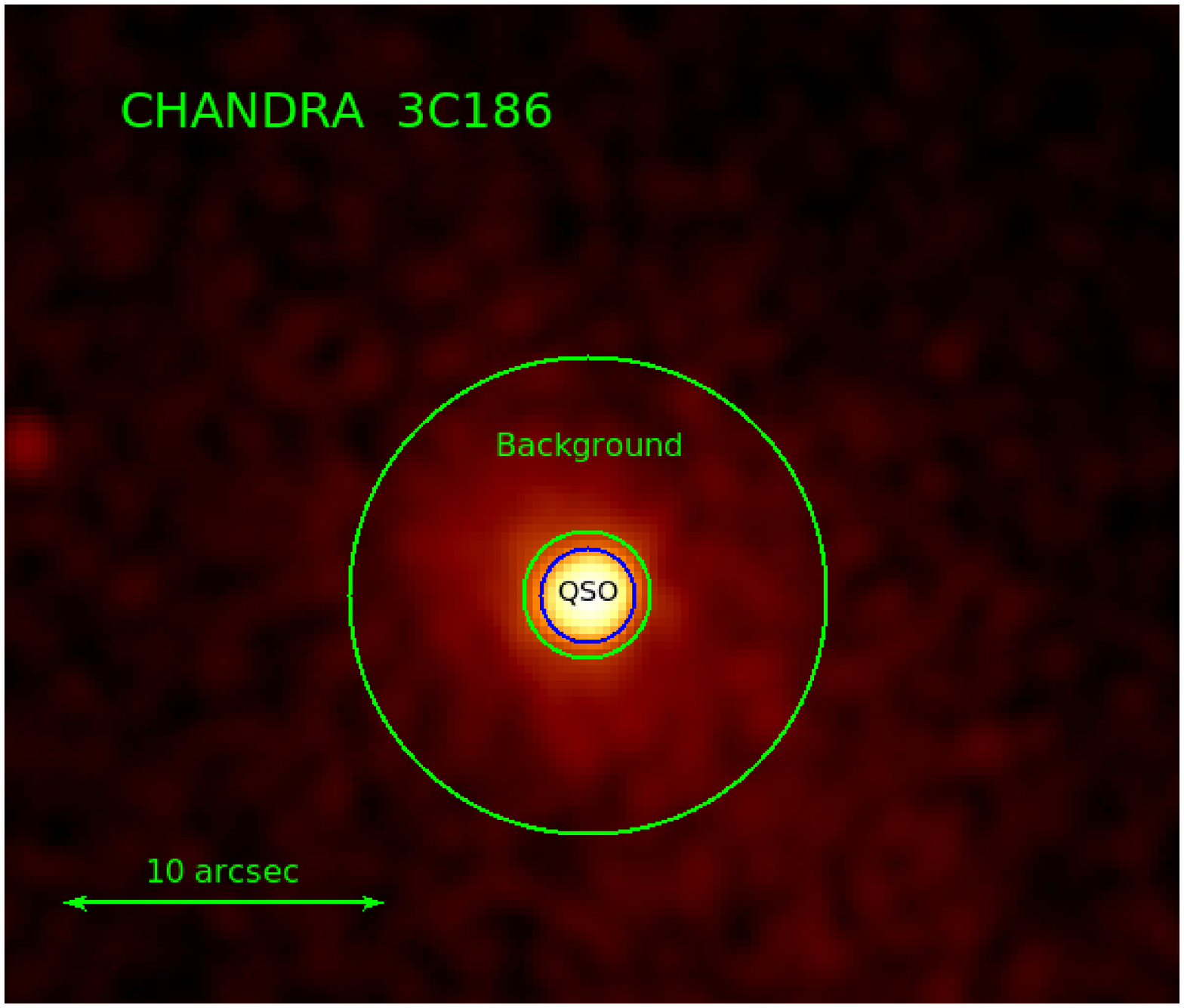}\hspace{1.0cm}
\caption{ Smoothed exposure corrected image of the  \Cha ACIS-S observation of 3C 186 in the 0.5-7 keV energy range. The regions selected for the quasar and background spectra are marked by a blue circle and a green annulus respectively.}

\label{f0}
\end{figure*}

\noindent
3C 186 has been observed for the second time with the \Cha Advanced
CCD Imaging Spectrometer \citep[]{Wei02} on 2007 December 03. The
$\sim198$ ksec long observation has been segmented in four intervals
(see Table \ref{t0}) due to \Cha observing constraints (Proposer's Observatory Guide [POG]).\\
The source was located on the ACIS-S backside-illuminated chip S3 
  and was offset by $\sim1'$ in Y coordinates from the default aim-point position to
  make sure that the cluster is not affected by a chip gap.
The observation was made in full-window mode and VFAINT mode,
which ensures a more efficient way of determining the background
events and cleaning background, especially at higher energies.\\
We performed the X-ray data analysis using CIAO version 4.3 and the
calibration files from the last CALDB release (4.4.2).  This version
includes the upgrade of the ACIS-S contamination file ({\it
  acisD1999-08-13contamN0006.fits}), which accounts for the temporal
degradation of the detector quantum efficiency due to materials
deposition on the ACIS chips or optical blocking filters.\\
We ran the \texttt{chandra\_repro} script available in CIAO 4.3 to
reprocess the data and apply the newest calibration. We used the
default options ``check\_vf=yes'' and ``pix\_adj=EDSER'' to obtain the
highest resolution image data.  We also investigated the lightcurve
and found no periods of significant background flares or quasar
variations.  
We also reprocessed the archival data from the first \Cha
observation performed on May-16-2002 and applied the newest calibration.
The 2002 observation was affected by a background flare and we used
the standard filtering of the data based on the background lightcurve.
The filtered quasar spectrum has a background rate of
$<0.0007$\,cts\,s$^{-1}$ in the quasar region and the effective
exposure time of 34.4\,ksec. \\
Figure \ref{f0} shows the \Cha ACIS-S image (2007 epoch) in the 0.5-7.0 keV band:
both the central quasar and the extended cluster emission are clearly
visible. The whole quasar structure is included in a circular region
of radius smaller than 2$^{\prime\prime}$. The diffuse cluster emission extends
to $\sim$500 kpc from the central quasar. Hereafter, we present the
analysis on the quasar, always indicated as 3C~186, while we refer to
\citet{Sie10} for the analysis of the X-ray data of the cluster.

\subsection{X-ray Spectral Analysis}

\begin{figure*}
\centering
\includegraphics[scale=0.35]{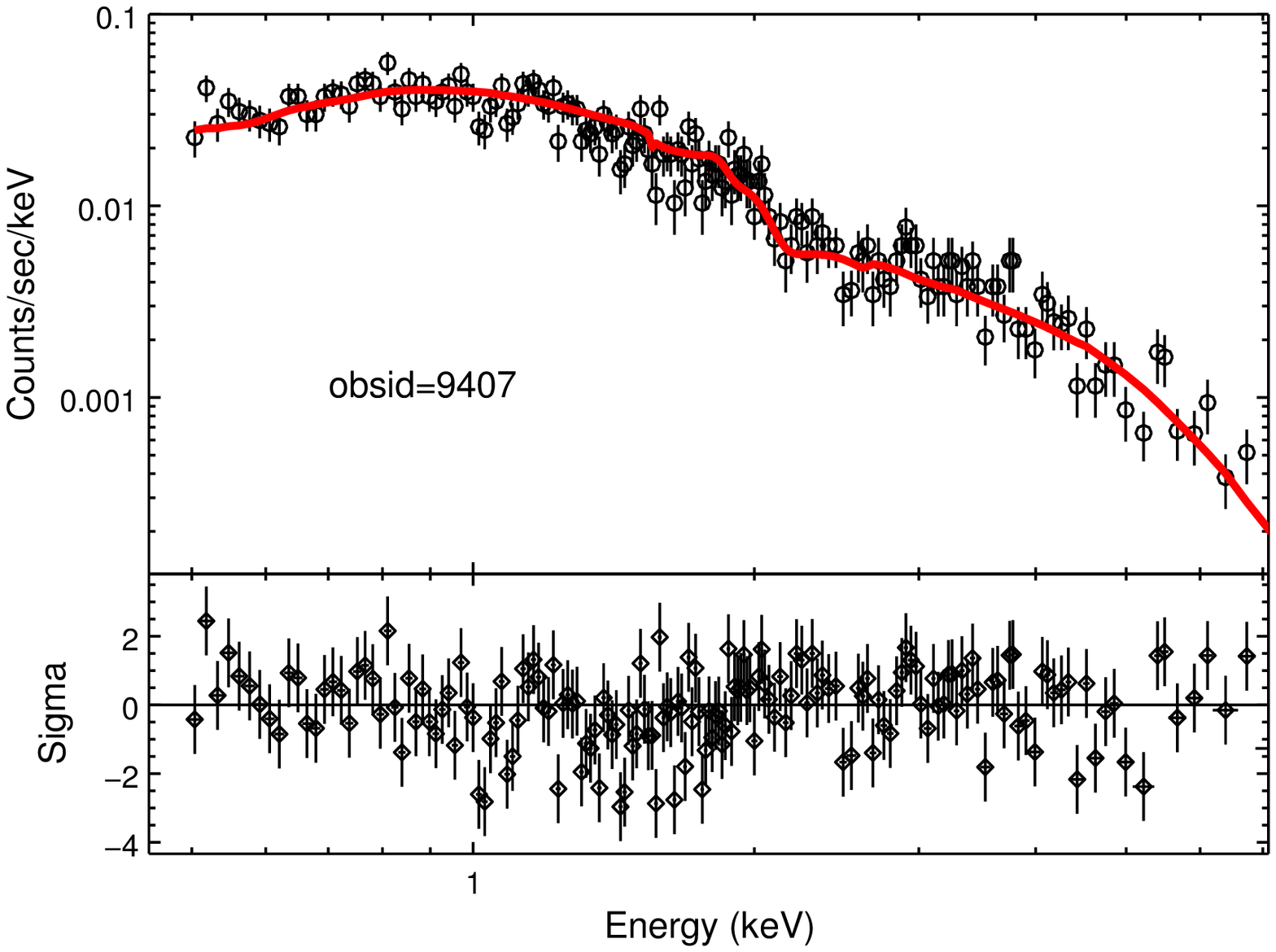}\hspace{0.1cm}
\includegraphics[scale=0.35]{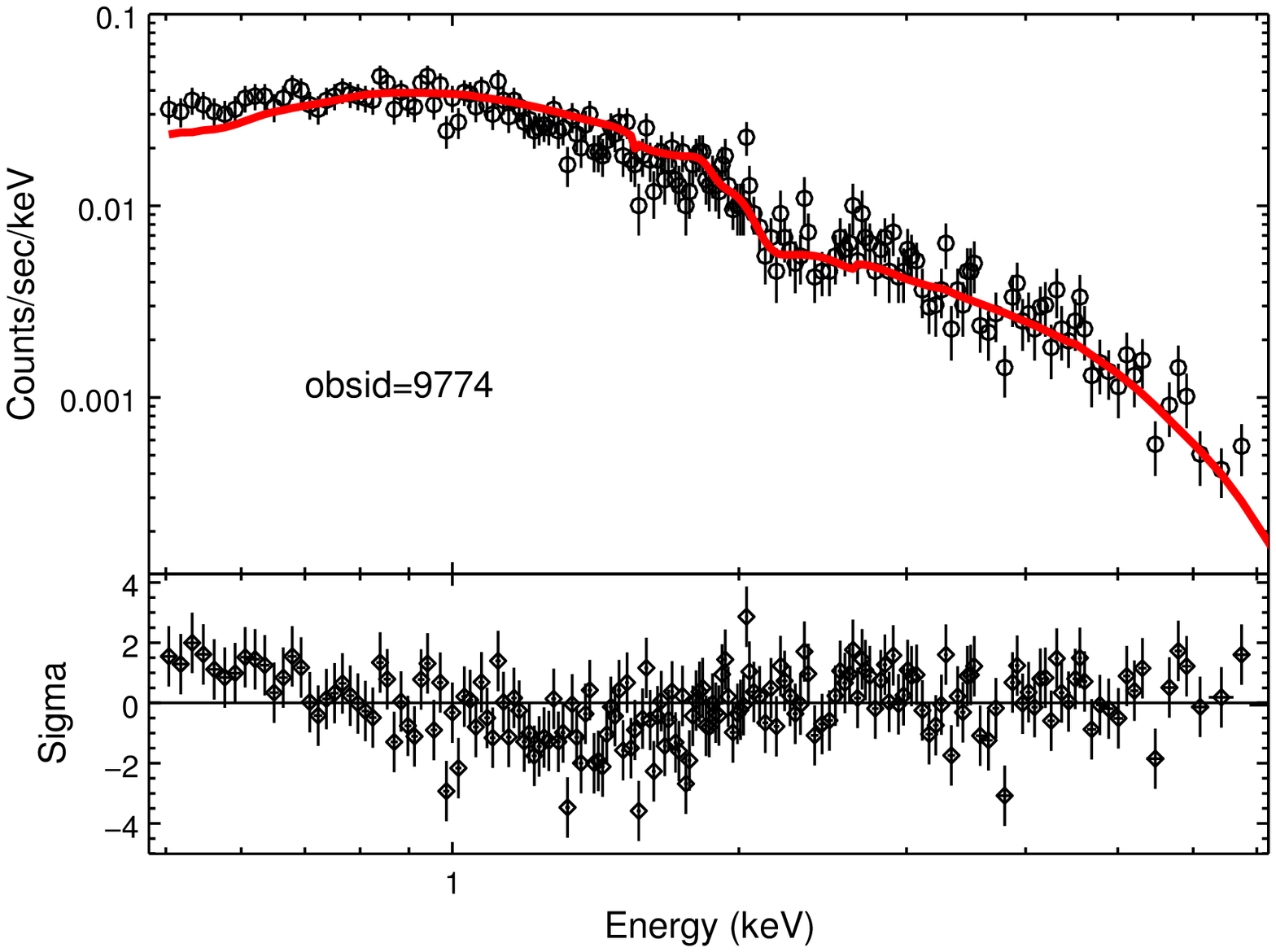}\vspace{0.1cm}
\includegraphics[scale=0.35]{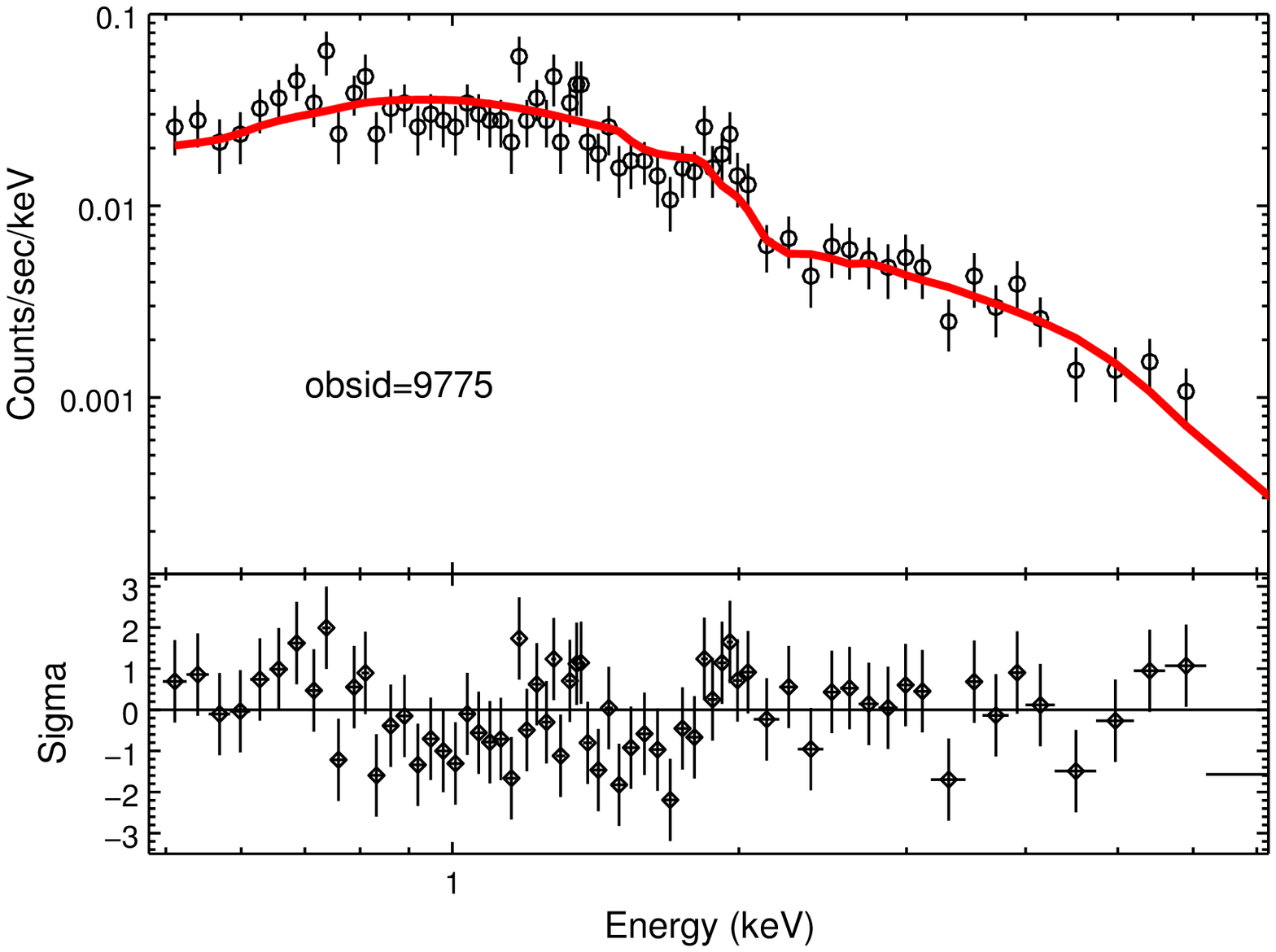}\hspace{0.1cm}
\includegraphics[scale=0.35]{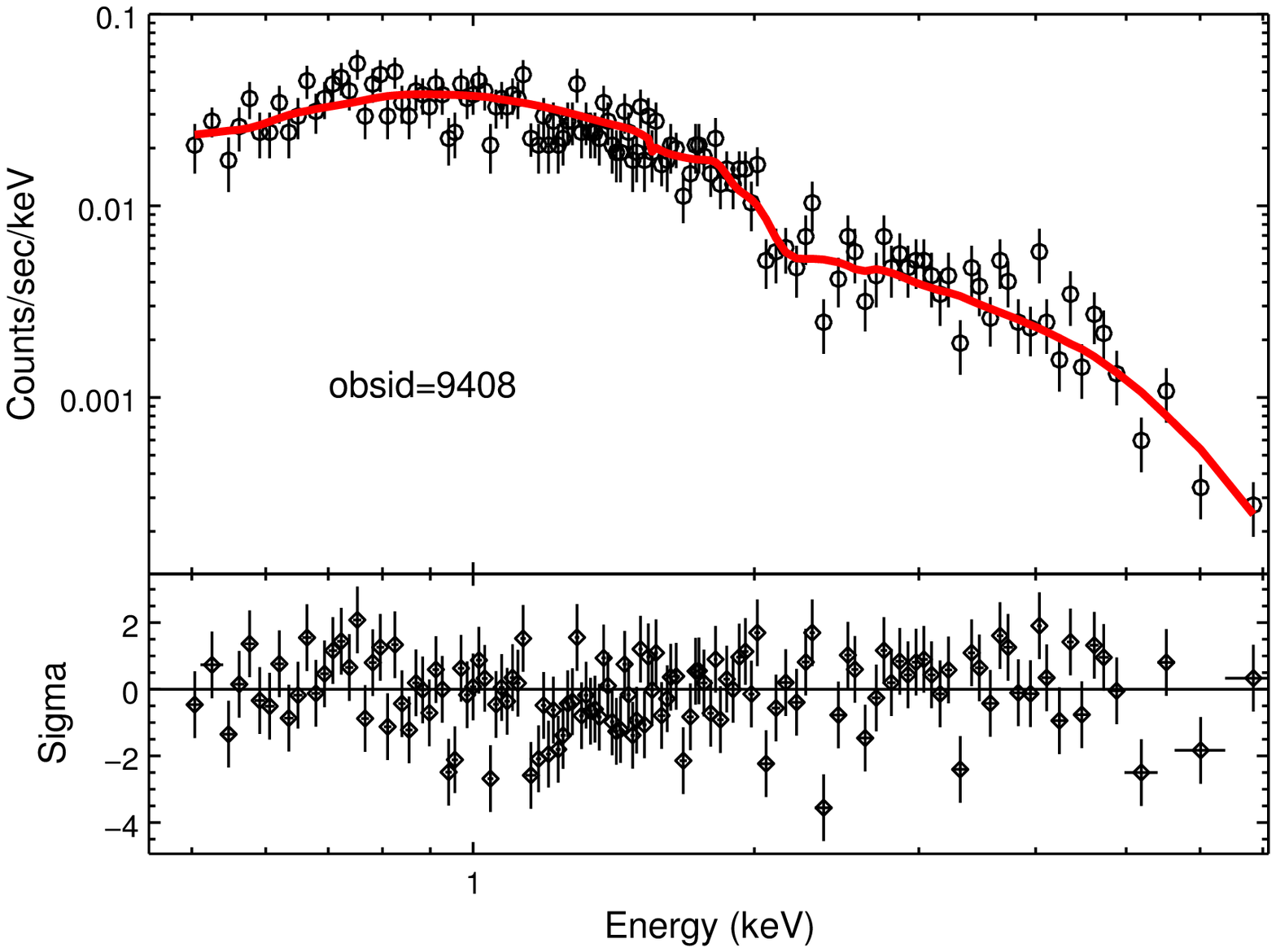}\vspace{0.1cm}
\includegraphics[scale=0.35]{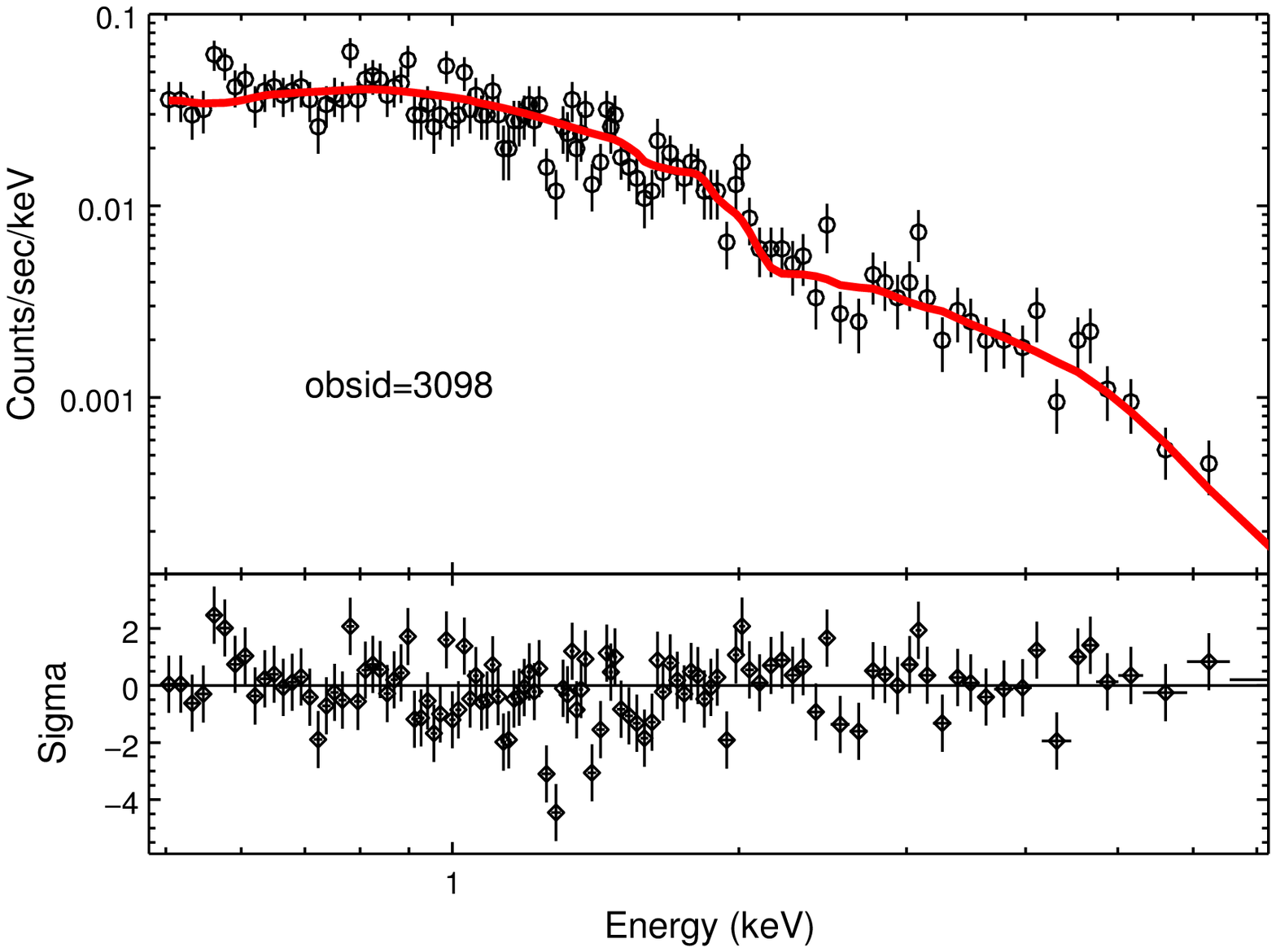}\vspace{0.1cm}
\caption{ACIS-S spectra in the 0.5--7.0 keV band of the five \Cha observations of the quasar 3C~186 with the best-fit absorbed power-law model superposed ({\it solid line}). The data were grouped to ensure a minimum signal-to-noise ratio of 3 per bin.  The lower panels show the residual difference between the model and the data in units of sigma. The inspection of the 9774 spectrum below 0.5 keV excludes additional source component at soft energies.}

\label{f1a}
\end{figure*}

The 3C\,186 radio source has a total angular size of
$\sim2^{\prime\prime}$.  Its complex radio structure is not resolved
out in the X-ray due to the \Cha angular resolution ($\sim1''$).  We
defined the quasar emission region as a circle, centered on the source
coordinates, with a radius r=1.5$^{\prime\prime}$, in order to collect
$98\%$ of the whole source emission.  An annulus surrounding the
quasar with inner and outer radius set at 2$^{\prime\prime}$ and
7.5$^{\prime\prime}$ was selected for the background.  In this way, a
source spectrum and the relative background dataset were obtained for
each of the five (1+4) observations. The total counts for each dataset
are shown in Table \ref{t0}.  Because of the increase of the
background at low and high energies,
we restricted the analysis to the 0.5--7.0 keV energy band.\\
We used Sherpa 4.3 \citep{2001SPIE.4477...76F} to model simultaneously
the four source spectra obtained in the individual observations during
2007 epoch. We used Cash statistics and the Nelder-Mead Simplex
optimization method \citep{NM65} to fit the models to the spectral
data. We also applied the parametric model to the background data and
used it in fitting the quasar spectra. All the statistical errors were
calculated at 90\% confidence
limits for a single parameter with the \texttt{conf} routine in Sherpa.\\
Because the thermal cluster emission dominates the background in the
vicinity of the quasar we selected an absorbed thermal plasma model as
a background model. We fit this model to the background data first and
then applied it with the appropriate scaling when fitting the quasar spectra.
We recall that the cluster contribution to the quasar spectrum is negligible
\citep{Sie10}. \\
We assumed an absorbed power law ({\it xsphabs} and {\it xspowerlaw}
models in Sherpa) model for the quasar emission.  The equivalent
column was fixed at the Galactic value $N_{H}=5.64\times10^{20}$
cm$^{-2}$, calculated using
COLDEN\footnote{http://cxc.harvard.edu/toolkit/colden.jsp}.  First we
fit the spectrum from each observation, then, as no significant
systematic difference was found, we fit the same power law model to
all four spectra simultaneously. The best fit power-law model
parameters are listed in Table \ref{t1}. In the 0.5-7~keV energy
range, $\Gamma_X=1.92\pm0.03$, while the photon index of the
individual observations ranges between 1.80$\pm0.08 - 1.94\pm0.03$
with the lower value for the data set with the worst S/N spectrum. An
inspection below 0.5 keV of the spectrum extracted from the 9774
observation made us exclude that the residuals we see in its 0.5-0.7
energy band are related to an additional emission component.
The unabsorbed fluxes between 0.5-2~keV and 2-10~keV, extrapolated
from the best-fit model, are respectively $F_{0.5-2\,keV}=(1.65\pm
0.04) \times10^{-13}$ erg cm$^{-2}$ s$^{-1}$ and
$F_{2-10\,keV}=(2.5\pm 0.1)\times10^{-13}$ erg cm$^{-2}$ s$^{-1}$.
\\
While the photon index is in a relatively good agreement with the
results presented in \citet{Sie05,Sie08} ($\Gamma_X=2.01\pm0.07$), we
note some level of discrepancy between the flux values
($F_{2-10\,keV}=1.7\times10^{-13}$ erg cm$^{-2}$ s$^{-1}$). We
therefore used the reprocessed archival data from 2002 to extract the
quasar spectrum in exactly the same way as in the case of the new 2007
observation.  We assumed an absorbed power law model, with the
equivalent column fixed to the Galactic value, and the same background
model in fitting this quasar spectrum. Our best fit parameter values
for the 2002 observation are consistent with the reported earlier
results, i.e. $\Gamma_X=2.03\pm0.07$ and
$F_{2-10\,keV}=1.8\pm0.1\times10^{-13}$ erg cm$^{-2}$ s$^{-1}$ (and
$F_{0.5-2\,keV}=1.36\pm0.05\times10^{-13}$ erg cm$^{-2}$ s$^{-1}$).
We conclude that the quasar's flux has increased by about 40\%
and the spectrum became slightly harder.  We note that the 0.5-7
  keV source count rates in the 2002 and 2007 observations are not so
  discrepant (0.049$\pm$0.001 counts/sec in the 2002 observation and between
  0.051$\pm$0.002 and 0.054$\pm$0.001 counts/sec in the 2007 ones) and the difference in
  the 2-10 keV flux is determined by the time-dependent calibration of
  ACIS-S and the flatter power-law index in the second epoch.  
\\
We checked and did not detect any significant absorbing column
intrinsic to the quasar, with a 3-$\sigma$ upper limit to the
equivalent column of hydrogen of $<2\times10^{20}$ cm$^{-2}$.
\\
We also investigated whether there are emission features in the
spectra and determined that there is no evidence for the presence of
an emission line in the spectra, the detection of which, at $\sim$3
keV, was only tentative in the first \Cha observation \citep{Sie05}.
In Figure \ref{f1a} we show the spectra from each observation
overplotted with the best fit power law model and the residuals. The
residuals are consistent with no emission line.

\section{3C~186: Modeling X-ray Emission and Broad-Band SED}\label{sect:Model}

Let us consider the indicators of non-thermal X-ray emission (see
Introduction) for the specific case of 3C~186.  3C~186 is a powerful
radio quasar with a high radio-loudness value, RL=3.7 \citep{Mil11}.
The source is also X-ray bright and, given the value of RL, the jet
component might dominate, or be important in the X-rays. On the other
hand, \citet{Sie08} have shown that its X-ray to optical luminosity
ratio appears to be smaller with respect to the typical RL quasars
\citep[see Figure 9 in][]{Sie08}.  The steep X-ray photon index
($\Gamma=1.92\pm0.03$), together with the value of the
optical-to-X-ray luminosity parameter ($\alpha_{OX}=1.74$) are more in
agreement with the average values found for RQ quasars, i.e.
$\Gamma_X=2.03\pm0.31$ and $\alpha_{OX}=1.49\pm0.19$ respectively
\citep{Kel07,Sob09}.
Note however that, albeit the limited statistics, the available  X-ray spectral analysis did not reveal the typical features of accretion related X-ray emission, most notably the Fe K$\alpha$ line. \\
Thus, while the radio features seem to suggest a relevant jet-related
X-ray emission,
the optical to X-ray data rather support a disk-corona scenario. \\
While both scenarios, accretion- and jet-related emission, do not exclude X-ray variability at the observed level, flux and spectral variability can be very useful to disentangle the different components in those sources where the two contributions are competing in the X-rays \citep[as for the case of 3C~273 and in Broad Line Radio Galaxies, see][]{GP04,GP07,Kat07,Sam09,Cha11}.\\
In 3C~186, the flux increases and there is a slight (although not
statistically significant) indication of spectral hardening.
A possibility is that the jet component is varying with respect to a steady accretion-flow continuum. Simultaneous X-ray and high-resolution radio observations would be needed to better investigate this hypothesis and we will discuss the radio-X-ray connection in Section 5.\\
The core dominance\footnote{The core dominance parameter is defined as
  $CD=\log(\frac{F_{c}}{(F_{tot}-F_{c})})$, where $F_{c/tot}$ are the
  core and total 5 GHz flux densities in the source rest frame
  \citep{SR79}.} (CD) of the radio source is another indicator of the
jet orientation, and thus of the importance of the beamed component.
\citet{FZ03} reported for 3C~186 a CD value of $-$1.66, lower than the
average value found in their quasar sample, $<CD>$=$-$0.53$\pm$0.92.
Given this CD, the quasar should be classified as a lobe-dominated
source\footnote{There are several definitions of core/lobe-dominated
  sources. Here, we define a radio source core/lobe dominated when its
  core monochromatic luminosity at 5 GHz is greater/less than half of
  the total radio luminosity.}, with a likely minor contribution of
the beamed-jet emission to the total X-ray flux. However, the CD
parameter estimate suffers of uncertainties related, for instance, to
the resolution of the radio map.  In addition to this, the CD
parameter may not be a suitable indicator of the jet axis orientation
in young radio sources. In fact, theoretical models predict that
during the first stages of the radio source evolution, the extended
structures, e.g. the lobes, are overpressured, and thus radio
over-luminous with respect to the lobes of giant radio sources
\citep{Beg99}.

\subsection{Jet SED Modeling}
\begin{figure*}
\centering
\includegraphics[scale=0.6]{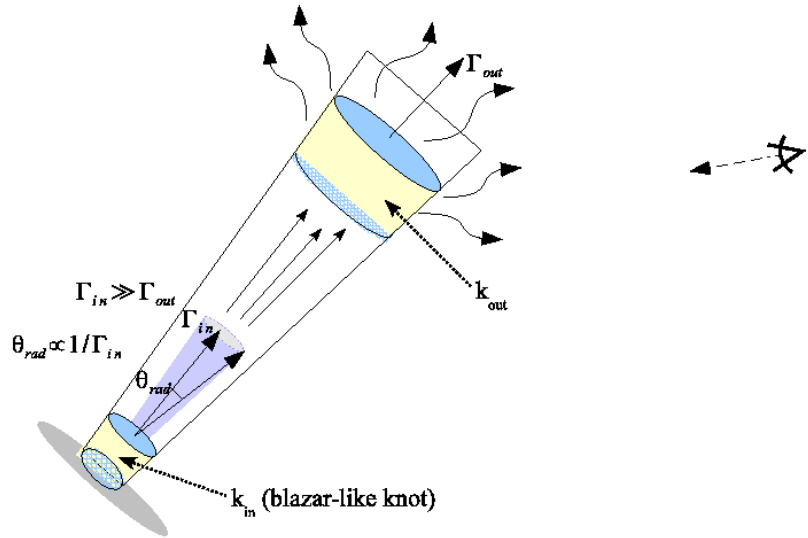}
\caption{Sketch of the jet structure for the case of an axial velocity gradient \citep[see][]{Cel01,GK03}.  We consider two regions of energy dissipation, a highly relativistic knot ($k_{in}$, blazar-like region) located at the base of the jet and a second, mildly relativistic one ($k_{out}$) farther out.
The synchrotron radiation from the blazar-like region is mostly emitted in a narrow cone ($\theta_{rad}\propto 1/\Gamma_{in}$, with $\Gamma_{in}$ bulk motion of the knot itself) and illuminates the second knot $k_{out}$.  The jet inclination with respect to the observer line of sight hides the bulk of the blazar-like emission to the observer view while the radiation from the slow-moving  knot $k_{out}$ is emitted quasi-isotropically.}  
\label{f2}
\end{figure*}

The modeling of the jet broadband SED can be effective in discerning
the X-ray emission components and possible jet contribution in RL
quasars.  This also provide us with an estimate of the jet kinetic
power (and the relative contributions of particles, magnetic field and radiation), a key quantity to investigate the dominant mode through which the radio source interacts with the surrounding medium. \\
We adopted a leptonic synchrotron and IC model to account for the
broad band emission of the 3C~186 jet.  The multi-wavelength dataset
allows us to constrain the most intense photon fields at different jet
scales and the available radio maps provide us with indications on the
sites where the bulk of the radiative dissipation is likely to occur
along the jet. The source is still embedded within the host galaxy
environment, therefore the jet is moving in a dense field of nuclear
photons, namely the optical--UV disk photons, the IR-torus photons and
possibly the IR-optical starlight photons. These components, pervading
the region where the jet energy dissipation occurs are relevant for
the jet IC emission and need to be
evaluated.  \\
In addition to this, we also investigated the possibility that the jet
has a complex structure, with
emitting regions moving at different velocities.\\
A structured jet has been invoked in order to efficiently produce high
energy (X-- to $\gamma$--ray) emission (see Celotti et al. 2001;
Stawarz \& Ostrowski 2002; Georganopoulos \& Kazanas 2003; Ghisellini,
Tavecchio \& Chiaberge 2005) as it has been shown that jets with a
velocity structure can be radiatively very efficient when a feedback
is established among the regions moving at different velocities.  A
stratified jet, with a fast spine and a slower sheath has been
considered to solve difficulties in models that unify BL Lac objects
and low-power radio galaxies \citep{Chia00}. Radio
\citep{Swa98,Gio99,Gir04} and X-ray \citep[see the case of Cen~A,
3C~273 and PKS~1127-145, ][respectively]{Wor08,Jes06,Sie07}
observations support the idea of a complex jet structure, although the
geometry is not still uniquely defined. On the basis of the results of
statistical studies, a velocity gradient seems also required in the
jets of powerful radio sources: in FR~II the bulk flow speed of the
jet significantly changes from parsec to kiloparsec scales
\citep{MH09}, and in RL quasars \citet{Mil11} infer different beaming
factors for the radio and
X-ray jet emission.\\
Models for  a structured jet typically consider a velocity gradient which develops along either the jet axis  \citep{Cel01,GK03}  or the cross-section radius  \citep{SO02,GTC05}. \\
Following \citet{Cel01}, we hypothesized a jet with two emitting
regions which are radiatively interacting: 1- a fast blazar-like
region close to the base of the jet; 2- a slower radiating knot
located farther out along the jet axis.\\
A sketch of the model is shown in Figure \ref{f2}. The blazar-like knot
is moving with a bulk Lorentz factor\footnote{the bulk Lorentz factor
  is defined as $\Gamma=(\sqrt{1-(v/c)^2})^{-1}$ where $v$ is the bulk
  velocity of the emitting plasma.}
 $\Gamma_{in}$ and emits via
the synchrotron mechanism. The synchrotron radiation from such
region is relativistically beamed within an aperture angle
$\theta_{rad}\propto1/\Gamma_{in}$ and illuminates the slower, outer knot ($k_{out}$). 
For $\Gamma_{in}>>1$, this results in an intense external field of seed synchrotron
photons in the $k_{out}$ frame. The bulk of the synchrotron -- and IC -- emission of the blazar-like component is however not visible by an observer whose line of sight is not closely aligned with the jet axis (at an angle $\theta>\theta_{rad}$).
Conversely, he/she could detect the synchrotron and IC radiation from
the slow-moving knot $k_{out}$, which is emitted more isotropically.
We considered this scenario for the case of 3C\,186
and discuss the conditions under which a structured jet can effectively reproduce the observed X-ray emission.

\subsubsection{Physical parameters}

In modeling the jet  broadband SED, there are some delicate points 
related to necessary assumptions on: the main parameters' values (the  magnetic field $B$ and
electron energy distribution, EED),
  the geometry of the emitting region, the spatial dependence of the local and external photon densities. 
The angle between the jet axis and the observer's line of sight $\theta$, and the bulk Lorentz factor $\Gamma$ are fundamental but also partly degenerate parameters with respect to the observables.
Multi-band data and radio maps allow to place some constraints on these quantities.\\

\noindent
{\it Jet radio morphology --} Radio observations provide us with spatial details on the 3C~186 jet
structure, necessary to constrain the physical parameters of the emitting regions. 
The radio map  at 1662 MHz in Spencer et al. (2001) (see Figure 2 in their paper  and available on NED\footnote{${\rm http://ned.ipac.caltech.edu/img11/1991MNRAS.250..225S/}$\\${\rm 3C\_186c:I:18cm:1991ssf.jpg}$})  shows a jet with two knots ($k2$ and $k3$) connected to the northern lobe $B$ by a radio low-luminous bridge.
In the southern lobe, $A$, observations at 600 MHz reveal the
presence of a hotspot \citep{Nan91}. The core, $k1$, is
self-absorbed below 1.6 GHz. The 600 MHz, 1.6, 5, 15 GHz measurements
are summarized in Table \ref{t2}.\\ 
The radio surface brightness of the knots ($6.62\times10^{-3}$
mJy/mas$^2$ for $k2$ and $1.96\times10^{-2}$ mJy/mas$^{2}$ for $k3$ at 1.6 GHz)
and hotspot ($0.67$ mJy/mas$^2$ at 600 MHz) is higher than that of
the diffuse structure ($1.79\times10^{-3}$ mJy/mas$^2$), supporting
the idea that the compact substructures are the main sites of power
dissipation. Therefore, in modeling the jet emission we 
considered only the two knots and the hotspot, and neglected the region of low radio brightness.\\
{\it Jet axis inclination and bulk motion --} Symmetry arguments
seem to rule out a very close alignment ($\theta\leq 10^{\circ}$) of the
source with the line of sight: the two lobes, $A$ and $B$, are
located at similar distances from the central region
\citep[respectively at 970 and 1250 mas, ][]{Spe91}.  On the other
hand, the detection of broad optical lines points to a dust free view
of the inner nuclear region. 
Thus, unless the  putative ``torus'' has a large opening angle,
the source does not lie close to the plane of the sky ($\theta\lesssim60^\circ$).  The visible
jet is pointing toward the lobe more distant from the core but is not on the same
side of the bright hotspot.  Estimates based on the one-sided VLA jet
give a $\theta\lesssim30^\circ$ angle and a corresponding de-projected size
of at least $\approx30$ kpc \citep{Sie10}.  
The value of the CD parameter (CD=$-$1.66), albeit the uncertainties and caveats, indicates a moderate alignment along the line of sight. 
Thus, we assumed an inclination angle of $30^{\circ}$ and discuss the
variation of the modeling results for smaller $\theta$.
\\
{\it Knot/hotspot physical parameters  --} We derived the deprojected distances from the radio core location of the two knots and the hotspot from the 18-cm radio data \citep[see Table 3 in][]{Spe91}. 
For the assumed $\theta=30^{\circ}$, these are $z_d(k2)\sim2.4$ kpc,
$z_d(k3)\sim9.8$ kpc and $z_d(hs)\sim15.4$ kpc for $k2$, $k3$ and the hotspot, respectively.
From these we calculated the physical
volumes using the 1.6 GHz maps for the two knots and the 600 MHz map
for the hotspot by assuming a cylindrical geometry (in Table
\ref{t2} the smallest
angular size corresponds to the base diameter and the longest to the
height).
\\
There is no information on the apparent motion of the jet features. The linear size to source age ratio provides us with some indications on $v_{ad}$ under the relevant assumption of continuous and uniform source expansion. 
For a linear size between the observed 15 kpc and  the deprojected (for $\theta=30^\circ$) 30 kpc, 
and an estimated source age of $\sim10^5$ yrs \citep{Mur99},  $v_{ad}$ ranges between 0.24$c$ and 0.49$c$. We note that hotspots' advance velocities of GPS sources are about 0.1--0.2$c$ \citep{Osw98,Gug05} while radio and statistical studies suggest mildly relativistic speeds ($\approx$0.5$c$ to $\approx$0.8$c$) of the jet at kiloparsec scales \citep{WA97,Har99,MH09,Mil11}.
Following \citet{MH09}, we assumed a bulk Lorentz factor of 1.4 for the
knots and adopted an upper limit of $v_{ad}=0.1c$ for the hotspot Lorentz velocity.
 
\noindent
{\it Magnetic field and EED --} The magnetic field, $B$,
and particle energy densities, $U_e$, were initially calculated under equipartition
assumption and normalized to the radio observed fluxes. 
As there are no evidences of a complex spectral shape from the observed SED, 
we assumed a simple power law for the
electron energy distribution, $N(\gamma)=K\gamma^{-p}$: $p$ is derived from
the lower limit of radio spectral index (i.e. $\alpha_r\leq 1.2$) $p=2\alpha_r+1=3.4$ and the
minimum and maximum Lorentz factors were set equal to $\gamma_{min}\sim50$
and $\gamma_{max}\sim10^5$. 
 The adopted $\alpha_r$ is steeper than the X-ray spectral index ($\alpha_X=\Gamma_X-1=0.92\pm0.03$). Nevertheless, the two spectral indexes are not necessarily expected to be the same if the X-ray flux is contributed by different competing components (e.g. the X-ray accretion continuum) and/or the X-ray emission is produced by the ``tails'' of the particle and seed photon distributions (as indeed the case here, see Fig. \ref{f4}).

\subsubsection{Photon fields}
In this section we evaluate the energy densities of the local and external photon fields which can act as target photons for the IC mechanism \citep[see ][]{Cel01}. 
As mentioned, we assumed that the bulk of the jet dissipation is spatially localized at the position of the 
radio compact features, namely the two knots and the hotspot. In the following, we inferred the energy densities of each photon field in the knot/hotspot reference frame (primed quantities) while the luminosities are the intrinsic ones in the reference frame of the source of photons ($L^i_{source}$). \\

\noindent
{\it Local fields --} The energy density of the synchrotron photons
produced locally in each knot and the hotspot, $U'_{SSC}$, was
calculated  
for a cylindrical geometry:
\begin{equation} 
U'_{SSC}\sim\frac{3 L^i_{syn}}{4\pi R h c}
\end{equation}
where $R$ is the cross section radius and $h$ the height of the cylinder (see Table \ref{t2}), and $L^i_{syn}$ the  integrated  synchrotron luminosity of the knots/hotspot estimated from the assumed EED and magnetic field ($L^i_{syn}(k2/k3/hot\,spot)=3.5\times10^{43}/9\times10^{43}/1.6\times10^{44}$ erg s$^{-1}$).\\

\noindent
{\it Nuclear fields, starlight and CMB --} 
The photon energy densities produced in the disk and torus as seen in the knot/hotspot frame ($U'_{disk}$ and $U'_{torus}$ for the disk and the torus, respectively)  depend on the distance $z_{d}$ of the knot/hotspot from the disk/torus and on the bulk motion of the knot/hotspot ($\Gamma_{(knot/hot~spot)}$). At the estimated distances of the jet knot/hotspot, the disk and the torus can be treated as point-like sources, and thus:
\begin{equation}
U'_{disk/torus}\sim\frac{L^i_{disk/torus}}{4\pi z_{d}^2 c \Gamma^2_{(knot/hotspot)}}
\end{equation}
where $L^i_{disk/torus}$ is the intrinsic luminosity of the disk/torus.\\
The estimated disk luminosity of 3C~186 is
$L^i_{disk}=5.7\times10^{46}$ erg s$^{-1}$ \citep{Sie05}. 
\\
The bulk of the IR luminosity of 3C~186 is likely to be thermal,  originating as disk radiation reprocessed in the dusty torus. Comparative studies of quasars and radio galaxies at redshift $z>1$ show that in quasars the heated-dust emission is typically a factor of 5-10 above the galaxy contribution \citep{Haa08}.
In order to account for the torus vertical dimension, we
assumed that all the dust emission comes from a height smaller than 10 pc
over the disk position. \\
The K-band magnitude of the 3C~186 host galaxy, $m_K(gal)=17.1$ \citep{Car98}, can be considered as a good indicator of the
starlight emission. The galaxy contribution corresponds to $\sim18\%$ of the total K-luminosity \citep{Car98}.  Here, we assumed that the most of the luminosity is produced within the core radius of the stellar distribution \citep[$\lesssim$1 kpc, see ][]{Rui05}. As the quasar emission dominates over the starlight in the IR and UV bands in the knot reference frame,  we neglected the starlight contribution.\\
In quasars jets, the observed X-ray emission from resolved knots can
be often consistently explained via Compton scattering of CMB photons
\citep{Tav00,Cel01}.  The energy density of the CMB photons
($U'_{CMB}$) in the jet frame is:
\begin{equation}
U'_{CMB}\sim aT^4_{CMB}(1+z)^4 \Gamma^2_{(knot/hotspot)}
\end{equation} 
where $T_{CMB}$ is the temperature of the CMB at $z=0$. This mechanism is efficient for highly relativistic plasma motion ($\Gamma\gtrsim10$), but requires also small $\theta$ in order to have the observed flux not relativistically de-boosted.\\
\citet{Cel01} have shown that in highly
relativistic sources the CMB radiation becomes the dominant field at $z_d\approx 30$ kpc for $z=0$. 3C~186 is at high redshift but has a small linear size, is likely not closely aligned to the observer line of sight and the assumed bulk motion is ``low''. Nevertheless, we included the CMB photon field in our analysis for comparison with the case of giant radio sources.\\
{\it External synchrotron photons --}  The energy density of the synchrotron radiation from the blazar-like component (see Figure \ref{f2}) in the reference frame of the knots  was estimated as \citep{Cel01}:
\begin{equation}
U'_{in}\sim \frac{L^i_{syn,bl.}\Gamma_{in}^4}{4\pi z_d^2 c\Gamma^2_{(knot/hotspot)}},
\end{equation}
where $L^i_{syn,bl.}$ is the intrinsic blazar-like synchrotron emission and the distance between the two radiating regions (i.e. the blazar-like and the knots) is approximated as $z_d$. The knots and hotspot correspond to the mildly relativistic knot $k_{out}$ in the sketch.  
We used the observed radio luminosity of the core to normalize $L^i_{syn,bl.}$.  Bulk Lorentz factor values between 10 and 20 are usually adopted to model the SED of blazar sources \citep{CG08,Ghi10}, with a few extreme cases of blazars in flaring state  \citep[for instance, in PKS 2155-304 the SSC modeling requires $\Gamma\approx 100$,][]{Fin08}. Here we assumed a moderate bulk motion of the blazar-like component, $\Gamma_{in}=10$, which gives an intrinsic  synchrotron luminosity $L^i_{syn}(k1)=8.5\times10^{44}$ erg
s$^{-1}$.\\  
Indeed, we do expect that the blazar-like component emits also via IC and the energy density of the blazar IC photons in the reference frame of the external knot can be calculated in the same way as $U'_{in}$, once we replace $L^i_{syn,bl.}$ with the intrinsic (blazar) IC luminosity. However, these high-energy photons are up-scattered to energies higher than the X-ray band we are considering here and for this reason we do not include them in our analysis.

\begin{figure*}
\centering
\includegraphics[scale=0.4]{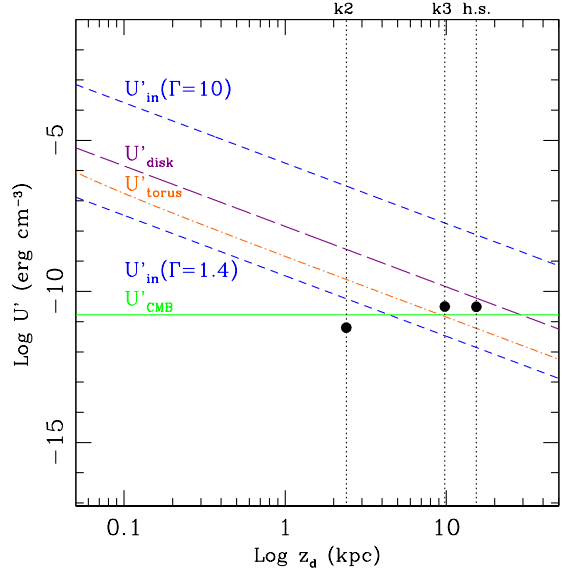}
\caption{Energy densities as a function of the distance from the base
  of the jet $z_d$ in the reference frame of the mildly relativistic
  ($\Gamma=1.4$) knot (see the sketch in Figure \ref{f2}). The
  external photon fields are: $U'_{disk}$ (long-dashed), $U'_{torus}$
  (dot-dashed), $U'_{CMB}$ (solid) and $U'_{in}$ (short-dashed). The
  energy density of the external synchrotron photons are estimated for
  a highly relativistic ($\Gamma=10$) knot (blazar-like) and, for
  comparison, for a mildly relativistic knot ($\Gamma_{in}=1.4$).
  The dotted vertical lines mark the deprojected position of the two knots and the
  hotspot in 3C~186 and the black  circles correspond to the
  energy densities of the local synchrotron fields ($U'_{SSC}$). The
  projected distance scale assumes $\theta = 30^\circ$.} 
\label{f3}
\end{figure*}

\begin{figure*}
\centering
\includegraphics[scale=0.5]{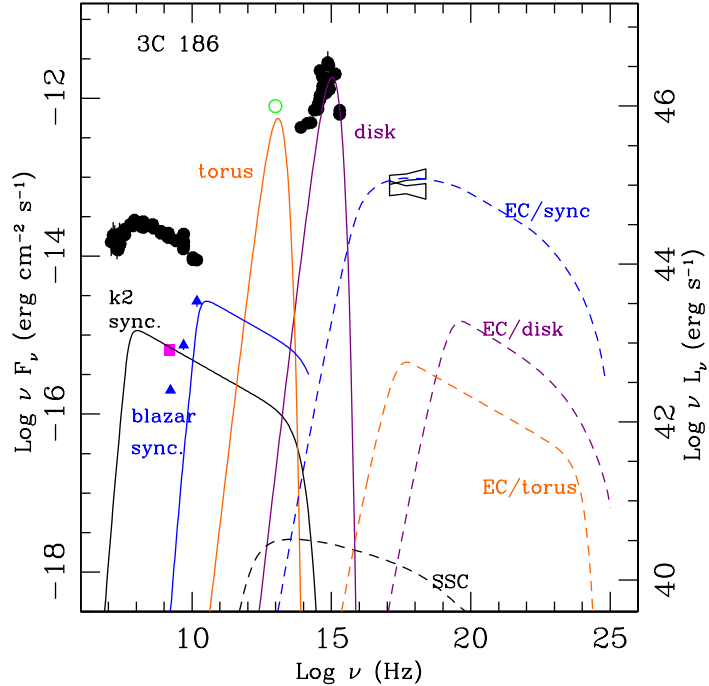}
\caption{Data and modeled SED of the knot $k2$, located at $\sim$2.4
  kpc distance from the jet apex. Data: black solid circles, green
  empty circle and the bow-ties are the radio to X-ray unresolved
  fluxes from the ASDC archive, Spitzer observatory and \Cha 2002 and 2007 observations (see text)
  respectively (bow ties show 90\% range). The magenta solid square is the 1.6 GHz flux for $k2$
  and   the blue solid triangles are the radio fluxes for the core, $k1$ \citep{Spe91}.   
Model: the black solid line (labelled k2 sync.) is the $k2$ synchrotron emission, the orange and violet solid lines are the torus and disk emission (labelled torus and disk respectively) and the blue solid line (labelled blazar sync.) is the synchrotron emission of the core, blazar-like, component (see text). The dashed lines correspond to the Comptonization of the synchrotron and nuclear photons with color correspondence with the seed photons as follows: SSC emission indicates the black dashed line, orange and violet dashed lines indicate the upscatterd torus (EC/torus) and disk (EC/disk) fluxes and the blue dashed line is the  EC of the external core synchrotron photons (EC/syn).}
\label{f4}
\end{figure*}

In Figure \ref{f3} we show the energy densities of the above photon fields
as a function of the distance $z_d$ for a jet (i.e. a knot in our
approximation) with a bulk motion  $\Gamma$=1.4.  The de-projected
positions of the two knots (assuming $\theta =30^\circ$)  and the
hotspot in 3C~186 jet are marked 
by vertical lines  and the values of the local synchrotron fields (i.e. seed photons for the synchrotron self-Compton) $U'_{SSC}$ are represented by solid points. \\
The blazar radiation field ($U'_{in}$) is the dominant one even at large
scales ($\approx 20$ kpc). For comparison, in addition to {\bf $\Gamma_{in}=10$} we also report the case in
which the jet has no velocity structure (i.e.{\bf $\Gamma_{in}=1.4$}): the
nuclear fields, $U'_{disk,torus}$, are the most intense at the scale of the
first knot $k2$ ($z_d\lesssim$3 kpc). Due to the intensity of the disk
radiation, $U'_{disk}$ can be higher of, or comparable with, the
local synchrotron emission $U'_{SSC}$ up to the hotspot distance ($z_d\approx15$ kpc). We
note that $U'_{CMB}$ is relatively less important below
$z_d\approx 20$ kpc as the amplification factor related to the
knot bulk motion is small ($\Gamma^2_{knot}\approx 2$).

\subsubsection{Multiwavelength mapping results}
The comparison of the photon fields shows that the nuclear photons  and the ``beamed'' external synchrotron radiation dominate at the scales of the first knot $k2$. In addition, $U'_{disk}$ and $U'_{in}$ are more intense than, or comparable with, $U'_{SSC}$ even at the hotspot distance.
This and the fact that the volumes and energy densities of electrons in the knots and hotspot are similar imply that the bulk of the IC emission should be dominated by the component closer to the nuclear region. Thus, the emission from $k2$ provides a reasonable model approximation to model the jet high-energy emission. In the model of the structured jet, $k2$ corresponds to the slow moving knot ($k_{out}$ in Figure \ref{f2}).\\
In Figure \ref{f4}, we show the results of the SED modeling of the
$k2$ knot emission together with the observed 3C~186 SED. 
The black points are
the multiwavelength data taken from the ASDC
archive\footnote{http://tools.asdc.asi.it/SED/}, the green one is the
Spitzer flux \citep[$F_{15\mu m(1+z)}=8.2\pm0.6$ mJy,][]{Lei10} and
 the bow-ties the 2 keV fluxes from the data of the 2007 (higher value)  and 2002 (lower value) \Cha observations\footnote{It is evident that our results do not change in any relevant way by considering the 2002 \Cha flux.}.
The synchrotron emission of $k2$ ({\it black solid line}) is
normalized to the 1.6 GHz flux ({\it magenta point}).  Disk ({\it
  violet solid line}) and torus ({\it orange solid line})
have been modeled as simple blackbody emission peaking at $\sim10^{15}$ Hz
and $\sim10^{13}$ Hz, respectively.
The synchrotron emission of the blazar-like component is normalized using the observed radio emission of the core ($k1$), assuming a spherical region of radius  $R_{in}\sim0.1$ pc, moving with {\bf $\Gamma_{in}=10$} in a magnetic field $B'_{in}\sim1$ G.
The dashed lines represent the Comptonization of the
non-thermal, local (SSC) and the external synchrotron (EC/syn) photons,
and of the thermal, disk (EC/disk) and torus (EC/torus), photons by
the relativistic electrons in the $k2$ knot.\\
In the high energy band (X-- to $\gamma$-rays), the SSC emission gives
a negligible contribution ($L_{SSC}\leq 10^{40}$ erg s$^{-1}$).
The luminosities of the EC/disk and EC/torus (in the observer rest frame) can reach a few $10^{43}$ erg s$^{-1}$, still below the observed X-ray emission.
The EC/torus
and EC/disk luminosities peak at different frequencies: the
up-scattered torus photons are mostly observed in the 2-10 keV band and
the bulk of the Comptonized UV photons above $10^{19}$ Hz. We note that the slope of EC/disk curve differs (in the high energy part) from the EC/torus one because in the first case the Compton scattering occurs in the Klein-Nishina regime.
Overall, the contribution of the inverse Compton emission off the nuclear (disk/torus) and local-synchrotron photon fields is not significant in the X-ray band.  We will discuss in Sec. 5 how this result is affected by our assumptions. 
 We also estimated the IC X-ray luminosity of the blazar-like component in the observer rest frame, considering external Compton on BLR  and IR photons \citep{SBR94,Bla00,Sik09}, and obtained values $\lesssim10^{44}$ erg s$^{-1}$.
However,  with the same knot parameters, in the framework of the structured jet model, the observed 2007 \Cha flux can be generated via up-scattering of the synchrotron photons from the blazar-like component.
 Note that for compact (GPS) radio sources the same mechanism could also efficiently produce X-ray emission if the beamed synchrotron photons are IC-scattered by the electrons in the lobes.\\

\section{Discussion}\label{sect:Discussion}
In 3C~186, we aimed at estimating the jet contribution to the total
X-ray emission by modeling its broadband SED.
The results of the modeling show that the IC X-ray emission of the jet does not provide a relevant contribution except when Compton scattering off the beamed synchrotron emission from a blazar-like knot is considered.\\
The radiative X-ray efficiency of a structured jet is related to the
presence of a significant velocity gradient between the two regions
(as $U'_{in}\propto\Gamma_{in}^4/\Gamma_{knot/hot spot}^2$).
Furthermore, this configuration accounts for the properties of
relatively misaligned sources: the strongly beamed emission of the
blazar component is mostly hidden to the observer view\footnote{ The
  observed (monochromatic) synchrotron/SSC luminosity goes as
  $\delta_{bl.}^{3+\alpha}$, where the Doppler factor is defined as
  $\delta=[\Gamma(1-\beta \cos \theta)]^{-1}$ and $\beta=v/c$, while
  the external Compton radiation follows the pattern
  $\propto\delta_{bl.}^{4+2\alpha}$ \citep{Der95} and, for the assumed
  values of $\theta$ and $\Gamma_{in}$ in 3C~186,
  $\delta_{bl.}\sim0.7$. }, while the radiation from the slow knot is
emitted quasi-isotropically ( $\delta_{k_{out}/k2}\sim1.8$ for the
mildly relativistic knot). At the optically thin radio frequencies
($\gtrsim$ 5 GHz), the beamed core-jet emission still dominates over
the quasi-isotropic synchrotron emission from the slower knot (Figure
\ref{f4}). This qualitatively agrees with the picture of a jet-related X-ray emission beamed to a lesser degree than the radio one \citep{Mil11}.\\
Before considering the implications in terms of the jet dynamics and
energetics in the jet structured model, we need to evaluate the robustness of these results. In the following, we discuss the main assumptions and verify if a different but ``reasonable'' set of parameter values could lead to significantly different conclusions.\\

\noindent
{\it Equipartition assumption --} 
Although equipartition (or minimum energy conditions) between magnetic field and particles are usually assumed in relativistic jets, hotspots and lobes, there are some indications that jet knots could be far from the minimum-power condition and be particle-dominated \citep{KS05}. If this were the case for the 3C~186 jet, the equipartition assumption leads to an underestimate of the IC fluxes. Nevertheless, for the one-zone jet, rather severe departures from equipartition are required to account for the observed X-ray flux: these span from $\sim$4 order of magnitudes for the SSC emission to a minimum of $\sim$ 2 order of magnitudes in the case of EC/torus emission.\\
\noindent
{\it Beaming factor ($\delta$($\Gamma$, $\theta$)) --} The viewing
angle $\theta$ and bulk Lorentz factor $\Gamma$ are parameters constrained together by the observables. A change of $\theta$ may determine a modification of the bulk motion or/and intrinsic luminosity.\\
In the structured jet scenario, if we assume a smaller value of $\theta$, say $\theta=10^\circ$, the blazar-like component becomes radio over-luminous with respect to the observed core flux, unless we assume lower values of $\Gamma_{in}$ or/and $L^i_{syn,bl}$.  This implies that $U'_{in}$ decreases and the final EC/syn with it (leaving unchanged $\Gamma_{out}$).\\
In the one-zone model, a change of $\theta$ is unimportant when the
jet/knot is moving with $\Gamma=1.4$.  At $\theta=10^\circ$, the
luminosity of the knot is maximally amplified for $\Gamma\sim 5$ (and
$\delta\sim6$). However, the same relativistic correction applies to
both the synchrotron and the SSC luminosities ($\propto\delta^{4}$),
so that the source parameters have to be coherently modified to agree
with the observed radio fluxes: a larger $\delta$ must correspond to a
decrease of the intrinsic quantities ($B$, $L^i_{syn}$ and thus
$U'_{SSC}$), implying that the SSC flux does not increase and still underpredicts the observed X-ray flux.\\
In the case of the external Compton process, the nuclear photon fields
will be severely deboosted in the frame of the outer knot moving with
$\Gamma=5$ (as $U'_{disk/torus}\propto1/\Gamma^2$).  Different is the
case of the isotropic CMB photon field. Since
$U'_{CMB}\propto\Gamma^2$, a large bulk motion increases the ratio
between the Compton and synchrotron but still low inclination angles
are required. For $\theta=10^\circ$ and $\delta\sim5.7$ (the
  combination of parameters which maximizes the ``observed'' IC/CMB flux at this
  angle), it is not possible to account for the observed X-ray flux
  via IC/CMB, without assuming a large ratio of the particle to
  magnetic field density ($\approx100$). In order to avoid the
  far-from-equipartition problem \citep[see][ for a review]{HK06},
  smaller angles, $\theta\lesssim 6^\circ$, not supported by the radio
  data, and the values of $\Gamma$ in the range between 10-30 are
  required.\\
\noindent
{\it Location of the emitting knot --} We identified the sites of the
jet X-ray emission with the compact radio features, knots and a hot
spot.  In blazars and quasars, the location of the region where the
bulk of the jet energy dissipation takes place is a matter of debate.
In blazars, $\gamma$-ray fast variability provides us with upper limits on the distance from the central black hole of about 0.1-0.3 pc, i.e. still inside the broad line region. However, this scenario is questioned. \citet{Sik09} proposed that the high-energy spectrum could be due to Compton scattering off thermal IR photons by  electrons   located at parsec-scales distances from the nucleus. This would also explain the lack of bulk-Compton and Klein-Nishina features in the blazar spectra.\\
In 3C~186, moving the dissipative knot to a smaller distance $z_d$
clearly determines an increase of the intensity of the
$U'_{disk,torus}$ (see Figure \ref{f3}). For a mildly relativistic
knot ({\bf $\Gamma=1.4$}), the EC/torus emission can in principle
reach the observed X-ray flux at $z_d\sim 100$ pc. However, this
estimate does not account for the decrease of the region volume
consequent to the smaller jet cross section. Higher EC/torus
luminosities can be obtained if the knot is moving within the torus
regions, as proposed for the X-ray emission of the quasar PKS
1127--145 \citep{Bla04}.  Nevertheless, this also require a small
$\theta$ and large $\Gamma$ \citep[in PKS 1127 -- 145 $\Gamma=10$ and
$\theta\sim 5^\circ$,][]{Bla04},
not supported by the 3C~186 data.

\subsection{Implications of a Structured Jet}
The above discussion on the main parameters confirms that it is hard
to accommodate the observed 3C~186 X-ray emission with a jet origin in
the framework of a single-zone scenario.
At the same time, the model which can successfully reproduce the observed X-ray emission in terms of jet emission implies a  structured jet. This has important consequences not only for the jet radiative efficiency but also for its dynamics and energetics.\\
\\
{\it Jet Dynamics --} A significant reduction in the velocity of the
jet with the increasing distance from the centre is the determinant
factor to ensure its radiative efficiency in the adopted scenario.  In
giant radio sources, the hotspots likely have sub-relativistic
velocities, however the jet/knot dynamics is still uncertain.  In FRII
sources, the detection of X-ray emission from knots is commonly
explained with IC/CMB mechanism \citep{Tav00,Cel01} and implies
relativistic jet velocities ($\Gamma\approx10$) on kpc scales.\\
On the other hand, radio observations support a significant
deceleration from parsec to kiloparsec scales, with characteristic jet
speeds at kpc distances in the range between 0.6$c$ and 0.7$c$
\citep{WA97}. \citet{MH09} adopted a Bayesian parameter-inference
method to constrain the jet Lorentz factors for a complete sample of
FRII with $z<1$ and found bulk motions of $\Gamma \approx$10-14 and
$\Gamma \approx$1.18-1.49 in the beams on parsec and kiloparsec
scales, respectively.
While the parsec-jet velocities agree with the ones inferred from VLBI observations \citep[][and references therein]{Lis09}, the results on the kpc jet challenge the beamed IC/CMB models. Jet deceleration is also proposed to reconcile the discrepancy between the estimated jet speeds at kpc-scales and the bulk motions required by the IC/CMB model (see \citealt{Har06}  for the case of lobe-dominated quasars and FRII radio galaxies, and \citealt{Mar11} for the study of a subsample of flat spectrum radio quasars observed by \Cha).\\
The mechanism which would be responsible for the jet deceleration is
still not clear. Entrainment of external interstellar medium seems
to be less important in powerful FRIIs and quasars, which display well-collimated jets on hundreds of kpc scales. Interactions with the surrounding medium are instead satisfactorily invoked to explain the presence of a multi-layer structure in the jets of FRI sources \citep[see the case of 3C~31,][]{Lai02a,Lai02b}. \\
In the spine-layer model \citep[e.g.][]{GTC05}, the two regions are
co-spatial. Thus, the radiative feedback takes place at the same jet
scales ($z_d$), differently from our case where the interactive
regions are located at kpc distance.  In the latter case, the
observations can partly resolve, and thus provide us with the
constraints on, both of the two regions \citep{GK03}.
In the spine-layer scenario the radio emission might be completely dominated by one of the two components, preventing us to constrain the model parameters of the other.\\
Finally, we note that a critical issue for the structured-jet model is
represented by the one-sideness of the radio source. The detection of
only one jet at radio frequencies is usually explained in terms of
beamed radiation and suggests that the jets are moving with
relativistic speeds. The crucial assumption here is that the radio
source is symmetric and the same physical conditions occur in the jet
and counter-jet. Nevertheless, several studies have shown that mildly
relativistic speeds are sufficient to account for the observed jet
sidedness \citep[$\beta\approx0.4-0.7$,][]{WA97,Har99,AL04,MH09}.
Given the assumed $\theta$, our adopted bulk motion is consistent
  with the minimum value ($\Gamma\gtrsim1.25$) necessary to reproduce
  the observed jet/counter-jet ratio lower limit\footnote{Here,
      we estimated the jet speed $\beta\gtrsim0.55$
      ($\Gamma\gtrsim1.25$) from Equation (A10) in \citet{UP95} for
      $\theta=30^\circ$ in the case of a moving isotropic source.
      Assuming a continuous jet, we obtain a slightly larger - but
      still consistent with our value- limit, $\beta\gtrsim0.7$
      ($\Gamma\gtrsim1.4$).}, $J\gtrsim 81$ (measured from the MERLIN
  5 GHz map at $\sim$0.05$^{\prime\prime}$ resolution presented in \citealt{Lud98}, C.C.Cheung, 2012,
  private communication).
\\
\noindent
We now consider the observed 3C~186 X-ray variability in the framework
of the structured jet. Blazars are typically variable at radio
wavelengths (with different intensities and timescales). If the
Unified Model \citep{UP95} is correct and the base of the jet of
3C~186 hosts a blazar region, its putative radio variability
also implies a change in $U'_{in}$ and thus in the observed EC/syn
flux.
The X-ray variability can be accounted for by a  corresponding variation (about a factor of 1.4) of the observed core radio flux (e.g. the radio flux of $k1$), that could be caused either by a fluctuation of the intrinsic luminosity $L'_{syn,bl.}$ , or by an increase of the bulk motion $\Gamma_{in}$. However, the propagation time of the blazar synchrotron photons to $k2$ introduces a significant delay ($\approx$ 8000 yrs) between the radio and X-ray flux variation and makes this hypothesis difficult to be tested. If the X-ray flux variation is instead due to a change in the distribution of the emitting electrons in $k2$, high resolution radio observations should be able to detect a simultaneous variation of the radio and X-ray flux.\\

\noindent
{\it Jet Power --}  An estimate of the total power carried by the jet can be inferred from the physical parameters adopted to model the two emitting regions of the structured jet.
The kinetic powers in particles and magnetic field were estimated as follows:
\begin{equation}
L_{i}=\pi R^2 \Gamma^2 \beta c U'_i
\end{equation}
where $U'_i$ is the energy density, in the comoving frame, of
electrons, $U'_e=m_e c^2 n_e<\gamma>$, cold protons, $U_p=m_p c^2 n_p$,
and magnetic field, $U'_B=B^2/8\pi$. Here, we assumed one cold proton
per electron ($n_p=n_e$), $m_e$ and $m_p$ are the electron and proton
masses, respectively, and $<\gamma>$ is the average electron Lorentz factor.\\ 
The analogous component associated to radiation, i.e. the radiatively dissipated power, is:
\begin{equation}
L_r=\pi R^2 \Gamma^2 c U_{rad}'\approx L' \Gamma^2
\end{equation}
where $L'$ is the  intrinsic total rest frame (synchrotron and IC) luminosity. \\
The energetics of the two emitting regions in the 3C~186 jet are shown in
Table \ref{t3}.  The values obtained for $k2$ are rather standard for
powerful sources, while the kinetic power corresponding to 
the blazar-component parameters ($L_{kin}=1.4\times10^{48}$ erg
s$^{-1}$) is at the high energy tail with respect to what is found
usually in blazar sources
\citep[ranging between $10^{46}$ and $10^{48}$ erg s$^{-1}$][]{CG08,Ghi09,Ghi10}. 
Indeed, the assumption on the jet composition affects in a critical way the value of $L_{kin}$.
In the  ``opposite'' case of a purely
electron-positron plasma, the jet would be magnetically dominated
($L_{B}=2.1\times10^{47}$ erg s$^{-1}$) and
highly dissipative ($L_r=1.1\times10^{47}$ erg s$^{-1}$).\\ 
The debate on the jet composition is still open. Based on X-ray observations, \citet{SM00} argued in favor of  a jet containing more $e^+e^-$ pairs than protons but dynamically still dominated by protons. Similar conclusions have been reached by \citet{Sta07}, for the case of the hotspots and jet of Cygnus A, and \citet{CG08}. 
Similarly, \citet{GT10} placed an upper limit of 10 pairs per proton ($n_e\lesssim10 n_p$) in the relativistic jets of FSRQs, in order to prevent radiation drag \citep[][and references therein]{Ode81,Sik96,GTC05}, when the dominant radiative process is external Compton.\\ 
The radiative power is a less model-dependent quantity, relying only on the assumption on the bulk motion and $\theta$, and provides us with lower limits of the jet power
of $\approx 10^{47}$ erg s$^{-1}$ and  $\approx 10^{45}$ erg s$^{-1}$ for the blazar-like component and the mildly relativistic knot, respectively.
\\
\cite{Sie10} estimated the jet power from the relation between the radio
luminosity at 151 MHz and the jet power itself \citep{Wil99} finding
$L_{kin}=2.4\times10^{45}$ erg s$^{-1}$. Considering the scatter in
the relation and the uncertainties on the composition, this is
not too far from the $L_{kin}$ calculated for the mildly relativistic
knot $k2$ (and consistent with the $L_{kin}$ lower limit provided by the radiative power). \\ 
The difference in $L_{kin}$ between the blazar component and the knot, about 2
orders of magnitude, points to a jet which dissipates (radiatively) much more efficiently, $\sim10\%$ of the total power, on the blazar scales. Nevertheless, if the above estimates are reliable, it remains unclear in which form the rest of kinetic energy is released. Alternatively, we might be underestimating the knot kinetic power as this refers to the radiating component only.
A further possibility is that we might be observing two subsequent outbursts of different intensity \citep{RB97,Cze09}.
 Support to the idea of intermittent activity in radio sources comes from the observations of radio relics and  radio sources with a double-double morphology \citep[i.e. with two aligned but unequally sized pairs of lobes,][]{Lar99,Sch00,Saik06,Jam07}.
 \\ 
The jet power of the blazar-like component is larger than, or at least comparable with,
the accretion disk luminosity \citep[$L'_{disk}\sim 10^{47}$ erg
s$^{-1}$][]{Sie10}.  Recent studies of blazars, based on high energy (X- to
$\gamma$-ray) observations, find that the jet power is proportional to
(but larger than) the disk luminosity \citep[][]{Ghi10}.
If this is the case for 3C~186, the radiative feedback between the source and its environment might be energetically driven by the jet rather than by the accretion power, depending on the modality in which the power is conveyed into the ambient medium.\\

\section{Summary}\label{sect:Summary}
We investigated the contribution of the jet emission to the total high-energy radiation observed in the compact, young and powerful quasar
3C~186. The results of the spectral analysis, based on a deep \Cha
observation, are ambiguous. The best-fit model, a single power-law
with a steep slope ($\Gamma=1.92\pm0.03$, and a 2-10 keV luminosity
$L_{2-10}=(1.15\pm0.04)\times10^{45}$ erg s$^{-1}$), is
compatible with non-thermal emission from the extended radio jets and
lobes as well as thermal emission related to the
central accretion.\\
In order to place quantitative constraints on the jet X-ray contribution,
we modeled its broadband emission with a leptonic synchrotron-IC
model.
The most relevant photon fields at the jet scales have been taken into
account as possible IC target photons. \\
The SED modeling shows that  in the framework of a single-zone model the jet emission is not relevant in the X-ray band (for the set of main physical parameter values favored by the observables).\\ 
Different results are obtained when the hypothesis of a single emitting zone is relaxed. We considered a jet with a velocity structure, exemplified by two emitting regions with different velocities, namely a blazar-like component and an external, mildly relativistic knot.\\
We find that an X-ray flux comparable with the \Cha observed one can be produced via Compton scattering off beamed synchrotron emission from a  blazar-like emitting region by the relativistic electrons in a knot located at kpc scales.
Hence, in 3C~186 the relevance of the jet as a source of high-energy radiation seems intimately related to its dynamical structure.
Indeed, this does not rule out  the possibility of a competing contribution of the disk-corona to the X-ray emission.
We note that the estimates of the X-ray emission related to the
accretion disk based on \citet{KB99} give $L_X\sim
L_{UV}/10^{1.5}\sim 2\times10^{45}$ erg s$^{-1}$, similar to the observed X-ray luminosity. 
 The detection of 3C~186 at $\gamma$-ray energies would be decisive to discriminate the nature of its X-ray emission. Unfortunately, the predicted flux of the model is more than 2 orders of magnitude below the detection limit of the {\it Fermi}-LAT \citep{Atw09}. In the framework of the structured jet, radio variability correlated to the observed X-ray one can be detected if it is caused by a change of the electron population of the mildly relativistic knot, but not 
if it is related to a variation of the blazar component.\\
There are some interesting aspects related to the structured jet scenario considered here.
Dynamically, the jet has to decelerate on kpc scales. It requires an initial high jet kinetic power ($L_{kin}\sim10^{48}$ erg
s$^{-1}$), that is comparable with those estimated for the most powerful
blazars. Unless time dependent, in which case one should expect to find sources where the kpc scale power exceeds that associated with the blazar component, the jet has to experience 
strong dissipation on kpc scales to account for the difference in the jet power estimated using the physical parameters of the two emitting regions. 
Depending on the assumptions on the composition, the initial jet power could be up to two orders of magnitude larger than  the disk luminosity. Therefore, in such a case, the
interactions, with the host-galaxy medium first and the cluster environment then, could be
dominated by the jet.\\
The study of 3C~186  is a pilot case for a broader investigation of radiation processes in young radio sources. 
In our future work we will extend the modeling presented  here to  other young radio sources observed in X-rays, in order to: a) determine wether there is a consistent behavior with respect to the possible jet contribution; b) ascertain wether there are candidates suitable for detection in the $\gamma$-ray band to definitely assess the origin of the high-energy emission in the early phases of the radio sources' growth.

\begin{acknowledgments}
We thank the anonymous referee for comments that significantly improved the manuscript.
The authors thank C.C.Cheung for providing radio measurements, M.Guainazzi, {\L}.Stawarz for useful discussions and T. Aldcroft for help in improving the text.
G.M. thanks P.Grandi,  M.Sobolewska and F.Massaro for valuable discussions and suggestions.
G.M. is grateful to L.Ostorero for insightful comments and great help in improving the manuscript.
This research is funded in part by NASA contract NAS8-39073. Partial support for this work was provided by Fermi grant NNX10AO60G. Partial
support for this work was provided by the National Aeronautics and Space
Administration through Chandra Award Number GO2-3148A and GO8-9125A,
GO0-11133X  issued by the Chandra X-Ray Observatory Center, which is
operated by the Smithsonian Astrophysical Observatory for and on behalf
of NASA under contract NAS8-39073. A.C. thanks Center for Astrophysics for hospitality.
\end{acknowledgments}

\newpage

\begin{table}
\caption{\Cha observations of 3C 186}
\label{t0}
\begin{center}
\medskip
\begin{tabular}{lcccc}
\hline
Obs. Date              &Obs ID                &Livetime         &Source Counts$^{(a)}$    &Source Counts$^{(b)}$\\
                               &                             &(s)               &(total)                            &(bkg. subtracted)\\
\\
\hline
 
2007-12-03              &9407                &66269                 &3630               &3595$\pm$66\\
2007-12-06              &9774                &75141                 &4037               &3996$\pm$70\\
2007-12-08              &9775                &15934                 &825                 &816$\pm$32\\
2007-12-11              &9408                &39623                 &2054               &2054$\pm$50\\
2002-05-16$^{(c)}$  &3098               &34436                  &1719               &1699$\pm$46\\
\\
\hline
\end{tabular}
\end{center}
$^a$ Total (source and background) counts within the energy range 0.5--7.0 keV in the selected circular region centered on the quasar  (r=1.5$^{\prime\prime}$).

$^b$ Background subtracted counts in the selected circular region centered on the quasar (r=1.5$^{\prime\prime}$).

$^c$ The 2002 \Cha observation was reprocessed applying the newest calibration. The same radius as for 2007 \Cha observations was assumed for the extraction region.
\end{table}

\begin{table}
\caption{Best-fit Power Law Model$^{(a)}$.}
\label{t1}
\begin{center}
\medskip
\begin{tabular}{lccccccccccl}
\hline
\\
obsid  & $\Gamma_X$ & Norm &  cstat \\
& & (10$^{-5}$ph~cm$^{-2}$~s$^{-1}$) & (442 d.o.f)$^{(b)}$ \\
\\
\hline
\\
9407 &  1.94${\pm0.04}$ & 8.69${\pm0.27}$ & 483.1 \\
9774 &  1.90${\pm0.04}$ & 8.39${\pm0.28}$ & 496.3 \\
9775 &  1.80${\pm0.10}$ & 7.83${\pm0.54}$ & 479.4 \\
9408 &  1.94${\pm0.06}$ & 8.21${\pm0.32}$ & 479.4 \\
\\
All$^{(c)}$  & 1.92${\pm0.03}$ & 8.42${\pm0.16}$ & 3274.46 (3548) \\
\\
\hline
\\
\multicolumn{4}{c}{Background Model} \\
\\
\hline
\\
obsid  & kT & Norm &  cstat \\
&(keV) & (10$^{-4}$ph~cm$^{-2}$~s$^{-1}$) & (442 d.o.f) \\
\\
\hline
\\
9407 & 5.90$^{+0.98}_{-0.84}$ & 1.89$^{+0.14}_{-0.12}$ & 395.7 \\
9408 & 5.8$^{+1.3}_{-1.0}$ & 2.02$^{+0.19}_{-0.16}$ & 364.8 \\
9774 & 4.78$^{+0.61}_{-0.52}$ & 2.11$^{+0.14}_{-0.13}$ & 326.7 \\
9775 & 5.89$^{+2.9}_{-1.4}$ & 1.88$^{+0.27}_{-0.26}$ & 266.0 \\
\\
All$^{(d)}$   &5.27$^{+0.7}_{-0.4}$   &2.01$^{+0.08}_{-0.09}$    &1358.9 (1774)\\
\\
\hline
\\
\\
\multicolumn{4}{c}{Power Law Model Parameters for the fixed Background Model$^{(e)}$}\\
\\
\hline
\\
obsid  & $\Gamma_X$ & Norm &cstat \\
& & (10$^{-5}$ph~cm$^{-2}$~s$^{-1}$)  & (442
d.o.f) \\
\\
\hline
\\
3098 &  2.03$\pm$0.07 & 7.33$^{+0.32}_{-0.31}$ & 468.9 \\
9407 &  1.94$^{+0.05}_{-0.06}$ & 8.67$^{+0.33}_{-0.30}$ & 483.0 \\
9774 &  1.90$^{+0.06}_{-0.05}$ & 8.36$^{+0.35}_{-0.24}$ & 496.3 \\
9775 &  1.82$^{+0.10}_{-0.13}$ & 7.97$^{+0.46}_{-0.77}$ & 447.8 \\
9408 &  1.93$^{+0.09}_{-0.06}$ & 8.07$^{+0.56}_{-0.24}$ & 479.9 \\
\\
\hline

\end{tabular}
\end{center}
$^a$ The equivalent Hydrogen column fixed at the Galactic value of 5.64$\times10^{20}$~cm$^{-2}$.
Uncertainties are 90\% for one significant parameter. The background model was first fit to all spectra and then 
fixed at the best fit parameter values when fitting individual observations.\\
$^b$ Degrees of freedom for single spectrum fit.\\
$^c$ Excluding obsid 3098.\\
$^d$ Best fit parameter values of the background model of the simultaneous fit of the 2007 \Cha observations.\\
$^e$ The background model parameters were fixed at the values of the simultaneous best-fit. \\

\end{table}

\begin{table}
\caption{Radio data for 3C 186 components.}
\label{t2}
\begin{center}
\medskip
\begin{tabular}{rrrrrrrr}
\hline
\
Regions                     &$F_{600} ^{(a)}$   &$\theta_{1}\times\theta_{2} ^{(a)}$    &$F_{1.6}^{(b)}$    &$\theta_{1}\times\theta_{2} ^{(b)}$   &$\alpha^{15}_{600}$    &$F_{5} ^{(c)}$ &$F_{15} ^{(d)}$\\
                                    &(mJy)            &(mas$^2$)                                   &(mJy)            &(mas$^{2}$)                               &                                          &(mJy)      &(mJy)\\
\hline
\\
A(hotspot+  &$525$  &$52\times15$         &$520$
&$350\times250$  &$\geq1.2$      &--       &-- \\
South lobe)  &$584$  &$203\times42$       &-- &--           &--             &$90$   &$20$(South East)\\
$k_{1}$(nucleus)     &abs.   &--               &$12$
&--             &$\sim 0.0$      &$15$   &$21$\\
$k_{2}$(first knot)  &--       &--                 &$40$
&$93\times65$    &$\geq1.2$      &--        &--\\
$k_{3}$(second knot) &--       &--                  &$80$
&$85\times48$    &$\geq1.2$      &--       &-- \\
N-jet($k_{3}$ to B)  &--       &--                 &$130$
&$725\times100$  &$\geq1.2$      &--        &-- \\
Jet(total)           &$385$  &$140\times44$      &--
&--              &--            &$93$    &$25$(Central)\\
B(North lobe)    &$315$     &$224\times83$  &$290$   &$460\times320$   &$\geq1.2$   &--     &$9$(North West)\\
\\
\hline
\end{tabular}
\end{center}
$^{(a)}$ 600 MHz fluxes and angular dimensions from \citet{Nan91}; $^{(b)}$ 1.6 GHz fluxes and angular dimensions from  \citet{Spe91};  $^{(c)}$ 5 GHz fluxes from \citet{Lud98};  $^{(d)}$ 15 GHz fluxes (C.C.Cheung, private communication).
\end{table}

\begin{table}
\caption{3C~186 Jet Powers for the structured jet model (see
  Figure \ref{f2}).}
\label{t3}  
\begin{center}
\medskip
\begin{tabular}{lcc}  
\hline
\\
Luminosities                       &\multicolumn{1}{c}{blazar-like component/}                               &\multicolumn{1}{c}{$k_{out}$/}\\
(erg s$^{-1}$)                     &\multicolumn{1}{c}{core}                     &\multicolumn{1}{c}{$k2$}\\
\\
\hline
\\
$L_e$      &$3.2\times10^{46}$    &$3.7\times10^{45}$\\
$L_p$      &$1.1\times10^{48}$    &$5.2\times10^{46}$\\
$L_B$      &$2.1\times10^{47}$    &$9.4\times10^{44}$\\
$L_{kin}$  &$1.13\times10^{48}$    &$5.6\times10^{46}$\\
$L_r$       &$1.1\times10^{47}$    &$1.0\times10^{45}$\\
\\
\hline
\end{tabular}
\end{center}
Powers associated to the bulk motion of (emitting) relativistic electrons ($L_e$), cold protons ($L_p$), Poynting flux ($L_B$). $L_{kin}$ is the total jet kinetic power and $L_r$ the radiatively emitted power.
\end{table}

\end{document}